# Internet of things-based (IoT) inventory monitoring refrigerator using arduino sensor network

Jessica Velasco[1], Leandro Alberto[2], Henrick Dave Ambatali[3], Marlon Canilang[4], Vincent Daria[5], Jerome Bryan Liwanag[6], Gilfred Allen Madrigal[7]

[1,2,3,4,5,6,7]Electronics Engineering Department, Technological University of the Philippines, Manila, Philippines
[1,7]Center for Engineering Design, Fabrication, and Innovation, College of Engineering, Technological University of the Philippines, Manila, Philippines

**ABSTRACT**

This study presents a system that combines a conventional refrigerator, microcontrollers and a smart phone to create an inventory monitoring that can monitor the stocks inside the refrigerator wirelessly by accessing an Android application. The developed refrigerator uses a sensor network system that is installed in a respective compartment inside the refrigerator. Each sensor will transmit data to the microcontrollers, such as Arduino Yun and Arduino Uno, which are interconnected by the I2C communications. All data and images will be processed to provide the user an Internet of Things application through the cloud-based website Temboo. Temboo will have access to send data to the Dropbox. A smartphone is connected to the Dropbox where all the data and images are stored. The user can monitor the stocks or contents of the refrigerator wirelessly using an Android Application.

*Corresponding Author:*

Gilfred Allen Madrigal
Electronics Engineering Department,
Technological University of the Philippines,
Ayala Boulevard, Ermita, Manila, Philippines.
Email: gilfredallen_madrigal@tup.edu.ph

## 1. INTRODUCTION

The rate of technological advancement is increasing with time, society is looking to create and develop easier ways to live and lengthen their lives. The internet is a massive source of information that millions of people use and depend on every day. A refrigerator is a popular household appliance that consists of a thermally insulated compartment and a heat pump (mechanical, electronic or chemical) that transfers heat from the inside of the fridge to its external environment so that the inside of the fridge is cooled to a temperature below the ambient temperature of the room.

Nowadays, with our advanced technology, smart refrigerator is being used to develop the use of appropriate storing of food. However, this device is not economically friendly because it is expensive. Thus, this study presents an Internet of Things-based smart refrigerator using sensor network and Arduino YUN Microcontroller that is suitable to any individual that usually spend more time at work and have difficulties monitoring their food. The proposed system monitors the stocks remaining and deficient remotely and at real-time. Also, it will notify the user the inventory update on his/her fridge through Internet.

The Internet of Things (IoT) is a system of interrelated computing devices, mechanical and digital machines, objects, animals or people that are provided with unique identifiers and the ability to transfer data over a network without requiring human-to-human or human-to-computer interaction [1-19]. A wireless sensor network is a group of specialized transducers with a communications infrastructure for monitoring and recording conditions at diverse locations.

There are previous studies made in relation with the development of smart refrigerators. In [20], the Intelligent Refrigerator utilized infrared (IR) emitters to detect the products and notify the owner about the status of the products inside the refrigerator through short messaging service (SMS) and the shop owner through the Ethernet network. In [21], the Smart Refrigerator uses weight sensor or load cell in measuring the weight of products inside the refrigerator. It notifies the user through Bluetooth of its status. In [22], the Smart Fridge used photodiode and Node MCU as the sensor to detect product and as microcontroller that will process data from the sensor, respectively. In [23], an overhead camera is used for the refrigerator that uses face recognition that will provide user access to the refrigerator and uses RFID to detect and track the products inside the medical internet refrigerator. Meanwhile, in [24], the Low-Cost Smart Refrigerator used light sensor to check if the refrigerator is open or closed, IR distance sensor and camera that will take pictures of the food inside the refrigerator. Image processing is implemented for the smart refrigerator in [25] to recognize the inventories and items inside the refrigerator but its recognition was executed using template matching causing a crash in its image processing feature. Next, in [26], a smart freezer called HighChest was developed for promoting energy efficiency and proper food storage. Lastly, in [27], a system for managing items for the solar-powered freezer was proposed to monitor perishable and non-perishable goods through GSM and Android application.

The general objective of the study is to develop an IoT-based inventory monitoring system for the refrigerator so that the user will be notified about the contents inside the refrigerator using an Arduino Sensor Network system. Specifically, it aims to: (1) develop an Arduino sketch that will process the data received from the sensor network system and will transmit the data gathered to the IP address of the host which will be the cloud storage of the data, and (2) to create an Android application which is synchronized to the IP address of the host which will notify the user about the contents inside the refrigerator.

The study will include 2 Arduino microcontrollers: Arduino Uno for the receiving of the data from the sensor network and the Arduino Yun for the receiving of image captured by the camera and performing the Internet of Things application. The sensor network will start off by sensing the data per compartment and will throw the data to the Arduino Uno. Then, the Arduino Uno is connected via I2C communication protocol to the Arduino Yun, so the data received by the Uno will be transferred to the Arduino Yun. After receiving all the data needed, the Arduino Yun will capture image using the camera and will upload the data received and the image captured directly to the Dropbox. The Android application will now then display the data and image on the Dropbox.

## 2. RESEARCH METHOD
### 2.1. Hardware Description

The hardware part of the research study is mainly composed of a refrigerator, main control box of circuitry boards designed for the sensor networks installed inside the fridge, Arduino Yun and Uno and an android phone.

#### 2.1.1. Refrigerator

The refrigerator used has the following characteristics: a.) Smart type inverter for better efficiency and cheaper electricity cost b.) No frost c.) Two-door fridge.

#### 2.1.2. Main Control Box of Circuitry Boards

The control box is the shelter of all the wires and power lines of the sensor network system and the microcontroller used by the study. The network of sensors is composed of 6 pcs of weight sensors installed in the freezer compartment and vegetable bin, 2 ultrasonic sensors for each lane of the egg tray, switches for the refrigerator bin and lastly the camera that is installed to capture image for the body of the fridge.

#### 2.1.3. Arduino Uno & Arduino Yun

In the study, the sensors are all connected to the Arduino Uno and all the output data gathered through all the sensors installed inside the refrigerator will be received by this Arduino board. Arduino Yun receives the image of the camera as output data. In the circuitry, the Arduino Yun serves as the master and Arduino Uno as the slave. more in general, Internet of Things projects. Arduino Yun has the onboard Ethernet and Wi-Fi interfaces that send and receive data through them making the Arduino Yun suitable for IoT-based projects [28]. Refrigerator and Block Diagram of the Control Box as shown in Figures 2 and 3. Wiring Diagram of the Freezer Compartment and Actual setup in the freezer compartment as shown in Figures 3 and 4.

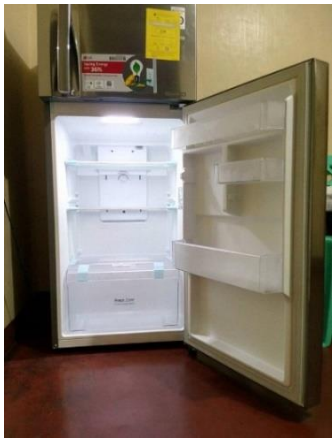

Figure 1. Refrigerator

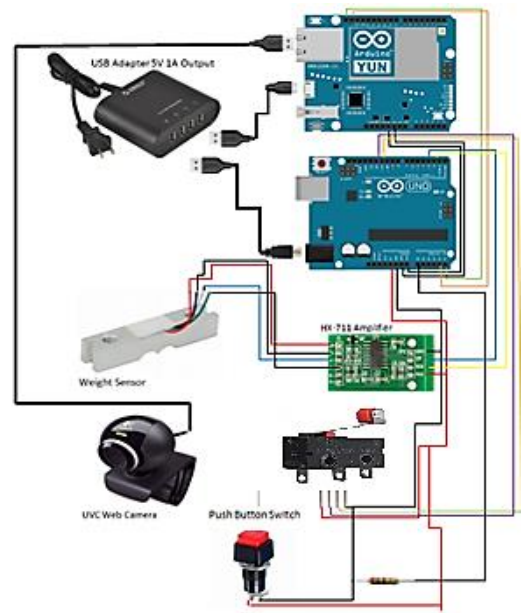

Figure 2. Block Diagram of the Control Box

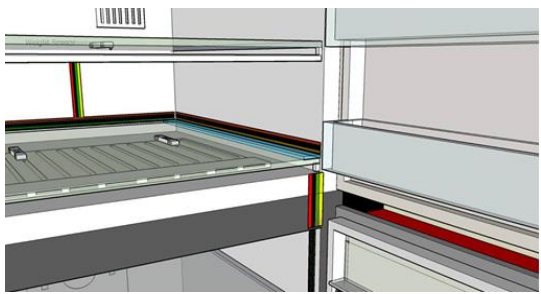

Figure 3. Wiring Diagram of the Freezer Compartment

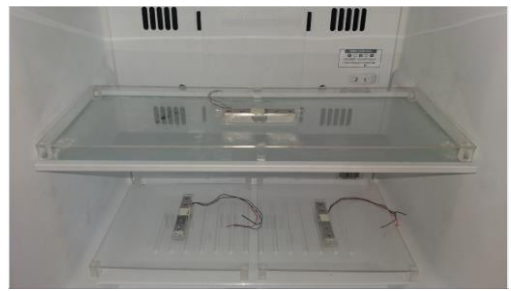

Figure 4. Actual setup in the freezer compartment

**2.2. Software Description**

The software of the study involves the language and codes used in the microcontroller, cloud storage, communication protocol of the 2 Arduino, interfacing website and the android application.

Figure 5 shows the basic flow of the whole process of the research study. The whole process basically starts with the refrigerator and shall proceed with the collection of data by the Arduino Uno using the sensor network and camera installed within it. The data captured will then be transferred to the Arduino Yun and will then be uploaded to the Dropbox which shall serve as the cloud storage of the research study. The Dropbox is linked with the Android application which shall display all the uploaded data on the Dropbox for the user to see it wherever the user is and whenever the user wants to see it.

**2.2.1. Arduino IDE**

The primary microcontroller of the project study is Arduino Yun which is connected to another microcontroller which is Arduino Uno. These 2 Arduino are connected through a serial communication known as I2C communication. Having 2 Arduino used in this project study, it is automatic that the group will have to use the Integrated Development Environment (IDE) as the main software of the project study.

**2.2.2. Android Application**

The Android application is developed using the Unity platform since it has a readily available application that allows Android application to receive data from any cloud storage devices. The Unity platform has a special application called BoxIt which can connect android application to the cloud storage devices which is in our case is the Dropbox.

### 2.2.3. Temboo

Temboo can connect microcontrollers and generate codes easily for such sensors to send data using internet. The group have registered an account in Temboo and will use this site as an interface for the IoT application of data from the Arduino Yun going to the Android Application made via the Unity platform. It will generate codes such ass token key, application key, token secret key and application secret key that are all important for the IoT application. It has a readily available code for its users and there are instructions that needs to be followed by the user for them to easily be able to transfer data through internet. It also has different access to different cloud storage devices such as Dropbox, Facebook, and other cloud storage devices. It uses choreos for uploading data to the net going to the Dropbox and this choreos are limited for those free accounts, just like in our group [28-30]. Flowchart of the study, Graphic user interface and Main program of the Android application as shown in Figures 5 till 7.

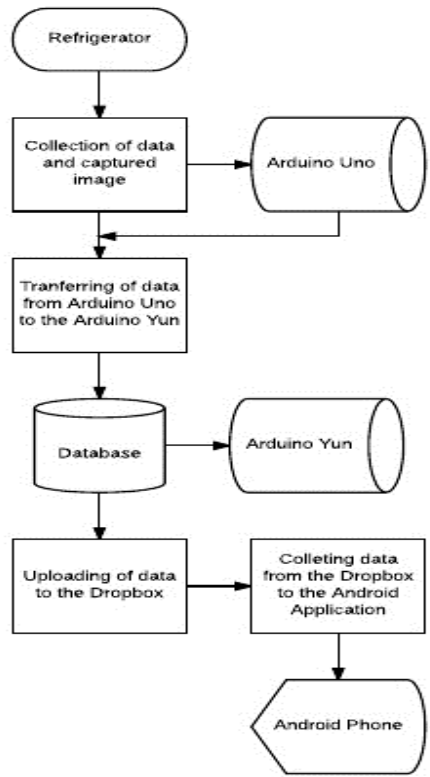

Figure 5. Flowchart of the study

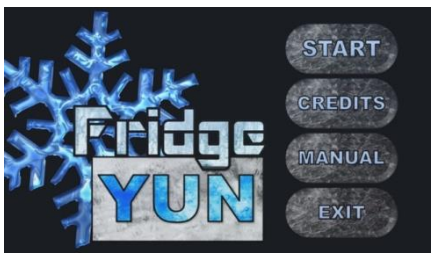   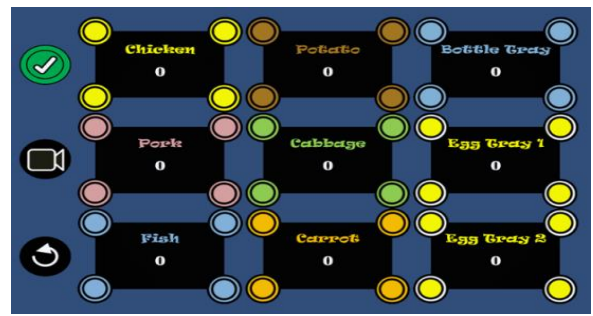

Figure 6. Graphic user interface    Figure 7. Main program of the Android application

### 2.3. Research Design

Figure 8 shows the system flowchart for the prototype. The sensors are placed accordingly in every compartment. The sensor will output whenever there is a presence or absence of an item. After the sensor outputs a value, the microcontroller will process the gathered data from the sensor which will transmit the data to the cloud storage using Internet of Things. Every time changes in the output of the sensor, the microcontroller will continuously transmit the data in the cloud storage. The smart phone will only receive the recent stored data in the cloud storage.

## 3. RESULTS AND ANALYSIS
The refrigerator interfaced with a sensor network system is divided into compartments such as freezer compartment and vegetable bin having weight sensors, limit switches for egg tray, push button switches for fridge bin and camera for the body.

### 3.1. Weight Sensors
Weight sensors are transducers that are similar to resistors in an electric circuit are affected minimally by temperature. So the group have gathered data during the off and on state of the refrigerator. The value of the pixel values of the weight sensors varies as the temperature of the fridge increases or decreases. The following are tables showing the realtionship of values of the temperature to the weight/force measured by the weight sensor.

Table 1 states an example of data gathered on the Weight Sensor #1. The data gathered on every weight sensor is different from the other weight sensor since the values are dependent on the HX 711 Amplifier. Meanwhile, Table 2 states the variation of values obtained on weight sensor #1 that has been observed by the researchers during the on state of the fridge in 5 different temperature levels. As you can see, the data is decreasing as the temperature on the fridge gets higher with some exemptions on the temp 4 due to certain reasons. Weight sensor inside the freezer compartment as shown in Figure 9.

### 3.2. Push Button Switches
Push button switches were used to determine the presence of bottled drinks on the refrigerator door below the eeg tray. The area was divided into four switches which means there will only be four bottled drinks that can occupy the area and it was divided based on the normal 1L-bottle size that is available in the market.

Table 3 shows on what level will a bottle drink be able to trigger the push button to detect the presence of the bottled drink. The data above is observed using a specific plastic container. Hence, the data above would vary depending on the what quality of material is the container made of.

### 3.3. Limit Switches
Limit switches were used to determine the presence of eggs on the egg tray compartment. They have softer triggering part and requires less force for them to be triggered as compared to push button switches. Every limit switch is allocated for every egg container im the egg compartment.

Table 4 states the limit switch triggering test shows the level or on what size and dimension of the egg would it be able to trigger the limit switch installed for the egg tray compartment. The trigger of the limit switch is dependent on the dimensions and quality of the egg. The good on the table means that the moment the egg was placed on the tray it has easily triggered the limit switch while the Loose on the table signifies that the egg did not easily triggered the switch due to certain reasons.

### 3.4. Camera
The camera was installed for the body of the refrigerator. The camera is installed on the center of the refrigerator door. The camera is an UVC (USB video class) type camera. Figure 11 shows the image captured by the camera installed inside the refrigerator and Limit switches inside the egg compartment as shown in Figure 10.

Table 1. Weight Sensor #1 at OFF State of Fridge

| Weight Sensor #1 | | | | | |
|---|---|---|---|---|---|
| Actual Weight (grams) | Measured Pixel Value | | | | |
| 0 | 183515 | 183606 | 183564 | 183476 | 183682 |
| 100 | 247793 | 247768 | 247768 | 247666 | 247848 |
| 200 | 313794 | 313888 | 313899 | 313869 | 313930 |
| 300 | 377682 | 377912 | 377798 | 377815 | 377904 |
| 400 | 444232 | 444350 | 444335 | 444353 | 444422 |
| 500 | 510112 | 510258 | 510134 | 510139 | 510106 |

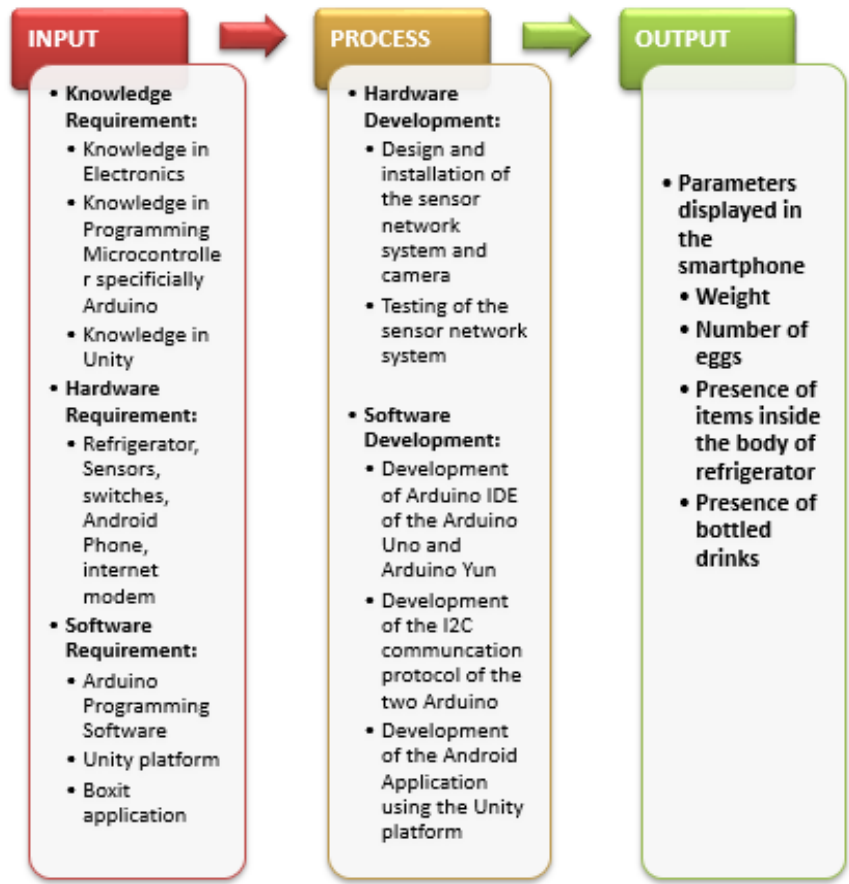

Figure 8. System flowchart

Table 2. Weight Sensor #1 at ON State of Fridge

| Weight Sensor #1 | | | | | |
|---|---|---|---|---|---|
| Actual Weight (g) | Temp 1 | Temp 2 | Temp 3 | Temp 4 | Temp 5 |
| 0 | 168423 | 159596 | 158966 | 161319 | 158627 |
| 100 | 221306 | 221306 | 221725 | 225132 | 223797 |
| 200 | 292241 | 289702 | 287881 | 289974 | 289642 |
| 300 | 358577 | 353506 | 355278 | 354455 | 355068 |
| 400 | 423648 | 421460 | 421402 | 421555 | 419959 |
| 500 | 494044 | 485257 | 484440 | 489729 | 484659 |

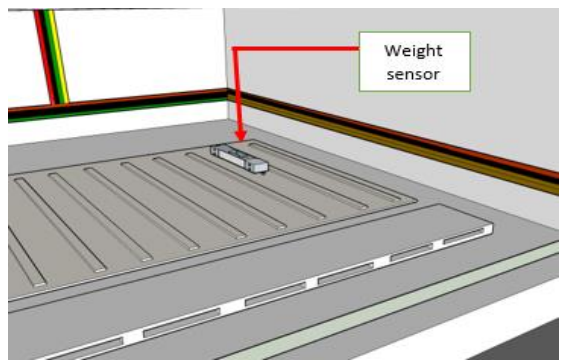

Figure 9. Weight sensor inside the freezer compartment

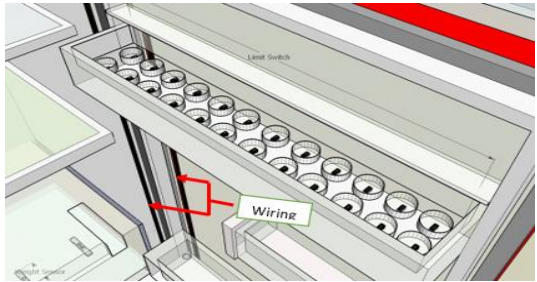

Figure 10. Limit switches inside the egg compartment

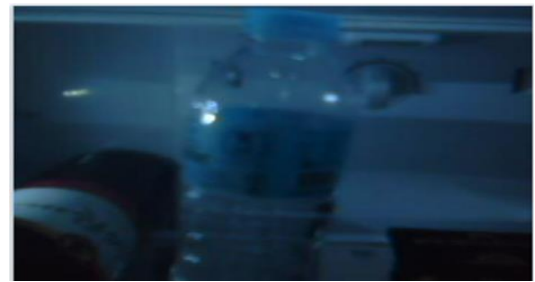

Figure 11. Image captured inside the refrigerator

Table 3. Push Button Trigger Testing

| Volume of the Bottled Drink | State of Push Button Switch |
|---|---|
| 0 ml | OFF |
| 100 ml | OFF |
| 200 ml | OFF |
| 300 ml | ON |
| 400 ml | ON |
| 500 ml | ON |
| 600 ml | ON |
| 700 ml | ON |
| 800 ml | ON |
| 900 ml | ON |
| 1 L | ON |

Table 4. Limit Switch Trigger Testing

| EGG TRAY | | | | | | | | |
|---|---|---|---|---|---|---|---|---|
| 5-peso eggs (small size eggs) | | | | | | | | |
| Tray No. | Switch 1 | Switch 2 | Switch 3 | Switch 4 | Switch 5 | Switch 6 | Switch 7 | Switch 8 |
| Tray No. 1 | Good | Good | Good | Loose | Good | Good | Loose | Good |
| Tray No. 2 | Good | Good | Good | Good | Loose | Loose | Good | Loose |
| 6-peso eggs (normal size eggs) | | | | | | | | |
| Tray No. | Switch 1 | Switch 2 | Switch 3 | Switch 4 | Switch 5 | Switch 6 | Switch 7 | Switch 8 |
| Tray No. 1 | Good | Good | Good | Good | Good | Good | Loose | Good |
| Tray No. 2 | Good | Good | Good | Good | Loose | Good | Good | Loose |
| 7-8 peso eggs (big size eggs) | | | | | | | | |
| Tray No. | Switch 1 | Switch 2 | Switch 3 | Switch 4 | Switch 5 | Switch 6 | Switch 7 | Switch 8 |
| Tray No. 1 | Good | Good | Good | Good | Good | Good | Loose | Good |
| Tray No. 2 | Good | Good | Good | Good | Good | Good | Good | Good |

## 4. CONCLUSION

An Internet of Things-based (IoT) inventory monitoring refrigerator has been successfully developed and was verified by certain people knowledgeable on the fields of IT as functional and efficient for the industry. The sensor network system inside the refrigerator has been successfully installed and has been working properly. All the wires were placed neatly, and the platforms of the sensors are positioned properly on their desired compartment. Furthermore, the wireless inventory monitoring system was a success and have been tried on different places using Android application developed using Unity and Boxit application.

Although the proposed system successfully operated, the following are needed to be improved or added it: (1) For a clear image of the body of the refrigerator, the camera to be installed must have a high rate of zoom out so that it can cover the entire body of the fridge; and (2) Distance-measuring sensors which can outstand the temperature of the refrigerator can be used for the egg tray compartment so that less wires would be required for the study.

## ACKNOWLEDGEMENTS

This study is supported by the University Research and Development Services Office and the University Research and Extension Council of the Technological University of the Philippines.

Indonesian J Elec Eng & Comp Sci   ISSN: 2502-4752   ❒   515

**REFERENCES**

[1] X. Bing, "Key Internet of Things Technology and Application Research," *Indonesian Journal of Electrical Engineering and Computer Science (IJEECS)*, vol. 12, no. 7, pp.5599-5602, 2014.
[2] A. ALFoudery, *et al.*, "Trash Basket Sensor Notification Using Arduino with Android Application*,*" *Indonesian Journal of Electrical Engineering and Computer Science (IJEECS)*, vol. 10, no. 1, pp.120-128, 2018.
[3] N. A. A. Bakar, Nur Azaliah Abu, *et al.*, "The internet of things in healthcare: an overview, challenges and model plan for security risks management process," *Indonesian Journal of Electrical Engineering and Computer Science (IJEECS)*, vol. 15, no. 1, pp. 414-420, 2019.
[4] Z. Kasiran, *et al.*, "An advance encryption standard cryptosystem in IoT transaction," *Indonesian Journal of Electrical Engineering and Computer Science (IJEECS)*, vol. 17, no. 3, pp. 1548-1554, 2020.
[5] A. G. Shabeeb, *et al.*, "Remote monitoring of a premature infants incubator," *Indonesian Journal of Electrical Engineering and Computer Science (IJEECS)*, vol. 17, no. 3, pp. 1232-1238, 2020.
[6] C.-P. Ooi, *et al.*, "FPGA-based embedded architecture for IoT home automation application," *Indonesian Journal of Electrical Engineering and Computer Science (IJEECS)*, vol. 14, no. 2, pp. 646-652, 2019.
[7] R. Ramly, *et al.*, "IoT recycle management system to support green city initiatives," *Indonesian Journal of Electrical Engineering and Computer Science (IJEECS)*, vol. 15, no. 2, pp. 1037-1045, 2019.
[8] C. Kim, *et al.*, "IoT task management system using control board," *Indonesian Journal of Electrical Engineering and Computer Science (IJEECS)*, vol. 13, no. 1, pp. 155-161, 2019.
[9] I. Kang, *et al.*, "User command acquisition based IoT automatic control system," *Indonesian Journal of Electrical Engineering and Computer Science (IJEECS)*, vol. 13, no. 1, pp. 307-312, 2019.
[10] T. S. Gunawan, *et al.*, "Performance Evaluation of Smart Home System using Internet of Things," *International Journal of Electrical and Computer Engineering (IJECE)*, vol. 8, no. 1, pp. 400-411, 2018.
[11] B. N. Fortaleza, *et al.*, "IoT-based Pico-Hydro Power Generation System Using Pelton Turbine," *Journal of Telecommunication, Electronic and Computer Engineering (JTEC)*, vol. 10, no. 1-4, pp. 189-192, 2018.
[12] R. T. M. Cruz, *et al.*, "IoT-based monitoring model for pre-cognitive impairment using pH level as analyte," *International Journal of Engineering Research and Technology*, vol. 12, no. 5, pp. 711-718, 2019.
[13] A. H. Abdulwahid, "Modern Application of Internet of Things in Healthcare System," *International Journal of Engineering Research and Technology*, vol. 12, no. 4, pp. 494-499, 2019.
[14] G. Mogos, "Critical Security Issues on Internet of Things," *International Journal of Engineering Research and Technology*, vol. 12, no. 1, pp. 113-118, 2019.
[15] R. J. Jorda, *et al.*, "Automated Smart Wick System-Based Microfarm Using Internet of Things," Lecture Notes on Research and Innovation in Computer Engineering and Computer Sciences, 2019, pp. 68-74.
[16] E. Galido, *et al.*, "Development of a Solar-powered Smart Aquaponics System through Internet of Things (IoT)", Lecture Notes on Research and Innovation in Computer Engineering and Computer Sciences, 2019, pp. 31-39.
[17] L. K. Tolentino, *et al.*, "Development of an IoT-based Aquaponics Monitoring and Correction System with Temperature-Controlled Greenhouse," 16th International SoC Design Conference (ISOCC 2019), in press.
[18] A. Thio-ac, *et al.*, "Development of a Secure and Private Electronic Procurement System based on Blockchain Implementation," *International Journal of Advanced Trends in Computer Science and Engineering*, vol. 8, no. 5, pp. 2626-2631, 2019.
[19] A. Thio-ac, *et al.*, "Blockchain-based System Evaluation: The Effectiveness of Blockchain on E-Procurements," *International Journal of Advanced Trends in Computer Science and Engineering*, vol. 8, no. 5, pp. 2673-2676, 2019.
[20] G. S. Nayak, and C. Puttamadappa, "Intelligent Refrigerator with monitoring capability through internet," *Int J Comput Appl*, vol. 2, pp.65-68, 2011.
[21] R. S. Khosla, *et al.*, "Smart Refrigerator*,*" *International Journal on Recent and Innovation Trends in Computing and Communication*, vol. 4, no. 1, pp.6-9, 2016.
[22] M. Edward, *et al.*, "*Smart fridge design using NodeMCU and home server based on Raspberry Pi 3.*," 2017 4th International Conference on New Media Studies (CONMEDIA), November 2017, pp. 148-151.
[23] V. Shadangi, and N. Jain, "*Medical internet refrigerator,*" 2015 International Conference on Control, Instrumentation, Communication and Computational Technologies (ICCICCT), December 2015, pp. 363-366.
[24] H. H. Wu, and Y. T. Chuang, "*Low-Cost Smart Refrigerator,*" 2017 IEEE International Conference on Edge Computing (EDGE), June 2017, pp. 228-231.
[25] J. K. P. Aranilla, *et al.*, "*Liveitup! 2 Smart Refrigerator: Improving Inventory Identification and Recognition,*" Research Congress 2013 De La Salle University Manila, 2013, pp. 1-9.
[26] M. Bonaccorsi, *et al.*, "'HighChest': An Augmented Freezer Designed for Smart Food Management and Promotion of Eco-Efficient Behaviour, *Sensors*, vol. 17, no. 6, pp. 1-21, 2017.
[27] J. J. A. Basa, *et al.*, "Smart Inventory Management System for Photovoltaic-Powered Freezer Using Wireless Sensor Network," *International Journal of Emerging Trends in Engineering Research*, vol. 7, no. 10, pp. 393-397, 2019.
[28] L. K. Tolentino, *et al.*, "*Aquadroid: an App For Aquaponics Control and Monitoring*," 6th International Conference on Civil Engineering (6th ICCE 2017), pp. 1-8, 2017.
[29] "Tools for Digital Transformation, Empowering Everyone to innovate with IoT, APIs & Emerging Tech", http://www.temboo.com/.
[30] A.P. Murdan and S. Caremben, "*An autonomous solar powered wireless monitoring and surveillance system,*" 2018 13th IEEE Conference on Industrial Electronics and Applications (ICIEA), May 2018, pp. 784-789.